\documentclass[11]{article}

\usepackage{amsmath}
\usepackage{amssymb}
\usepackage{amsthm}
\usepackage{graphicx}

\newtheorem{theorem}{Theorem}
\newtheorem{lemma}[theorem]{Lemma}
\newtheorem{proposition}[theorem]{Proposition}
\newtheorem{conjecture}[theorem]{Conjecture}
\newtheorem{corollary}[theorem]{Corollary}

\renewcommand{\vec}[1]{\mathbf{#1}}
\newcommand{\mat}[1]{\mathbf{#1}}

\newcommand{\Csym}{{C_{sym}}}
\newcommand{\Cpf}{{C_{PF}}}
\newcommand{\Cmm}{{C_{MM}}}
\newcommand{\Csum}{{C_{sum}}}

\title{On the Fairness of Rate Allocation in Gaussian
Multiple Access Channel and Broadcast Channel}

\author{\\ \Large Kenneth W. Shum and Chi Wan Sung \\
Department of Electronic Engineering \\
City University of Hong Kong \\
Email: {\tt \small kshum2006@gmail.com}, {\tt \small
itcwsung@cityu.edu.hk}}

\date{Draft, Nov 3, 2006}

\begin{document}
\maketitle

\begin{abstract}
The capacity region of a channel consists of all achievable rate
vectors. Picking a particular point in the capacity region is
synonymous with rate allocation. The issue of fairness in rate
allocation is addressed in this paper. We review several notions of
fairness, including max-min fairness, proportional fairness and Nash
bargaining solution. Their efficiencies for general multiuser
channels are discussed. We apply these ideas to the Gaussian
multiple access  channel (MAC) and the Gaussian broadcast channel
(BC). We show that in the Gaussian MAC, max-min fairness and
proportional fairness coincide. For both Gaussian MAC and BC, we
devise efficient algorithms that locate the fair point in the
capacity region. Some elementary properties of fair rate allocations
are proved.
\end{abstract}

\section{Introduction}

There are several fairness criteria in rate allocation. The simplest
one is to mandate that all users have the same data rate and
maximize this common data rate within the capacity region. This
allocation is equitable and maximize the data rate of the worst
user. However, it often does not effectively utilize system
resources.

Another criterion is called {\em max-min fairness}. It relaxes
equity and allows increasing the rates of some users without
lowering the minimum data rate in the system. Under such an
allocation policy, nobody can be benefited by worsening anybody who
has lower data rate. It can be rephrased as follows. If we take some
resources from a wealthy user to a relatively poor user (without
reversing the order of wealthiness), the resulting allocation is
considered fairer. Such operation is called a Robinhood operation.
We say that an allocation is max-min fair if no Robinhood operation
is possible without violating feasibility.

The max-min fairness is considered quite stringent. Any decrease of
rate of a user with low data rate cannot be compensated by
increasing the rate of any user with higher rate, no matter how
small the decrease is. Kelly considers logarithmic utility function
and proposes proportional fairness for network flow
control~\cite{Kelly97}. Roughly speaking, we say that a rate
allocation is {\em proportionally fair} if any adjustment will
decrease the sum of percentage change over all users. A framework of
optimization is thereby introduced using Lagrangian technique. Both
max-min and proportional fairness are popular criteria for flow
control in unicast and multicast networks~\cite{KMT98,
SarkarTassiulas02}.

Another classical notion of fairness is the {\em Nash bargaining
solution} coined by Nash in the 50's~\cite{Nash50}. In fact,
proportional fairness is its special case. The Nash bargaining
solution is a standard tool in cooperative game theory, and is
applied widely in network resource allocation. For example, see
\cite{HJL05} for application to orthogonal frequency division
multiple-access networks.

Yet there is another fairness criterion based on the theory of {\em
majorization}. We will give formal definitions of all the above
criteria in the next section.

In this paper, the problem of picking a point in the capacity region
of Gaussian MAC and BC according to some fairness criteria is
considered. Some works have been done on Gaussian MAC. The sum rate
at the max-min fair allocation is characterized
in~\cite{KapurVaranasi04}. The polymatroid structure of the capacity
region of Gaussian MAC is exploited in~\cite{MMK06, ShumSung06}, and
algorithms for locating the max-min fair point are found.
Implementation issue is addressed in~\cite{ShiFriedman05}.

The organization of this paper is as follows. In
Section~\ref{sec:definitions}, we give precise definitions of
fairness, and review the theory of majorization and Schur-convexity.
The results are summarized in Section~\ref{results}, and the details
for the Gaussian MAC and BC are in Section~\ref{sec:MAC}
and~\ref{sec:BC} respectively. The appendix contains some proofs of
the theorems in Section~\ref{sec:MAC}.

\section{Fairness, Majorization and Schur-convexity}
\label{sec:definitions} The set of all users is denoted by $\Omega =
\{1,\ldots, K\}$.

In this section, we review several fairness criteria, and the theory
of majorization. We will use the symbol $\mathcal{R}$ to represent
capacity region, which is assumed to be a closed and convex set
throughput this section.

\bigskip

An allocation is {\em symmetric} if every user has the same data
rate. {\em Symmetric capacity} is the maximal sum rate of all
symmetric allocations,
\[ \Csym(\mathcal{R}) := \max \{ Kr:\, (r,r,\ldots, r) \in \mathcal{R}
\}.
\]

An allocation is called {\em max-min fair} if we cannot increase the
rate $r_i$ of user $i$ without decreasing $r_j$ for some $r_j \leq
r_i$, while maintaining feasibility. At the max-min fair allocation,
no user can increase the data rate without compromising users with
lower data rate. Formally speaking, a rate allocation $\vec{r}^{MM}$
is max-min fair in $\mathcal{R}$ if for any $\vec{r}\in\mathcal{R}$
such that $r_i^{MM}< r_i$ for some $i$, then we can find
$j\in\Omega$ such that $r_j < r_j^{MM}\leq r_i^{MM}$. The sum of
rate at the max-min fair allocation for capacity region
$\mathcal{R}$ is called the {\em max-min capacity} and is denoted by
$\Cmm(\mathcal{R})$.

{\em Proportional fair} (PF) rate allocation $(r_i^{PF})_{i=1\ldots
K}$ is the data rate allocation that maximizes
\[\sum_{i=1}^K \log
r_i.\] The {\em proportional fair capacity} is the corresponding sum
rate,
\[
 \Cpf(\mathcal{R}) := \sum_{i=1}^K r_i^{PF}.
\]
Since the capacity region $\mathcal{R}$ is closed and convex, and
the $\log$ function is concave, the maximization is well-defined.
Another characterization of the proportional fair allocation
$\vec{r}^{PF}$ is
\[
 \sum_{i=1}^K \frac{r_i-r_i^{PF}}{r_i^{PF}} \leq 0
\]
for all point $\vec{r}$ in the capacity region $\mathcal{R}$.

\begin{figure}
\begin{center}
\includegraphics[width=60mm]{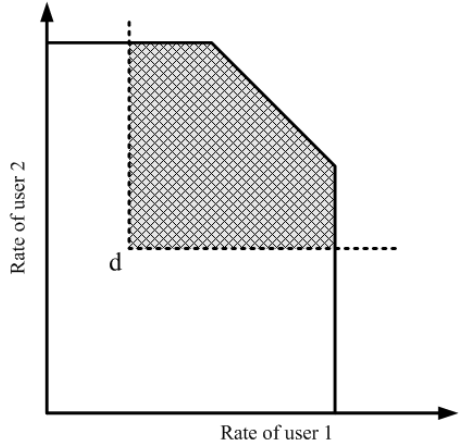}
\end{center}
\caption{The point $d$ is the disagreement point. The shaded area is
the acceptable rate allocation.} \label{fig:DP}
\end{figure}

In the {\em Nash bargaining solution}, there is a notion of {\em
disagreement point}, which is the default operating point if the
users fail to reach any agreement. User $i$ will not accept any data
rate lower than $d_i$. The rate allocated to user $i$ in the Nash
bargaining solution should be larger than or equal to $d_i$
(Fig.~\ref{fig:DP}). If we are given a disagreement point $\vec{d}$
in the capacity region, the Nash bargaining solution maximizes
\[
 \sum_{i=1}^K \log( r_i- d_i)
\]
over all points in the region
\[
 \{ \vec{r}\in \mathcal{R}:\, r_i \geq d_i \ \forall i\}.
\]
The Nash bargaining solution satisfies several desirable properties.
See~\cite{OR90} for details. It is obviously identical to the
proportional fair solution when the origin is chosen as the
disagreement point.

For a vector $\vec{x}=(x_1,\ldots x_K)\in \mathbb{R}_+^K$, we denote
the components in nondecreasing order by
\[
 x_{[1]} \leq x_{[2]} \leq \ldots \leq x_{[K]}.
\]
We say that vector $\vec{x}$ is {\em majorized} by vector $\vec{y}$,
written as $\vec{x}\preceq \vec{y}$, if for $k=1,\ldots, K-1$,
\begin{align*}
  \sum_{i=1}^k x_{[i]} &\geq \sum_{i=1}^k y_{[i]}, \text{ and} \\
  \sum_{i=1}^K x_{[i]} &= \sum_{i=1}^K y_{[i]}.
\end{align*}
In other words, if we sort the components of $\vec{x}$ and $\vec{y}$
in nondecreasing order, the cumulative sum of the components of
$\vec{x}$ is larger than the corresponding cumulative sum of
$\vec{y}$.

Majorization induces a partial order that measures dispersion. It is
known as the Lorenz order in economics, and is used for comparing
income distributions. When $\vec{x} \preceq \vec{y}$, we say that
the distribution according to $\vec{x}$ is less spread out, and is
thus fairer than that of $\vec{y}$. A canonical example is that the
vector
\[
 \left( \frac{1}{K}, \ldots , \frac{1}{K} \right)
\]
is majorized by any vector in $\mathbb{R}_+^K$ whose components sum
to 1.

A function $f: \mathbb{R}_+^K \rightarrow \mathbb{R}$ is called {\em
Schur convex} if
\[
 f(\vec{x}) \leq f(\vec{y}), \text{whenever } \vec{x} \preceq
 \vec{y}.
\]
If the inequality above is reversed, then we say that the function
$f$ is {\em Schur concave}. A class of Schur-convex functions is
constructed using the following lemma.

\begin{lemma}
If $\theta: \mathbb{R} \rightarrow \mathbb{R}$ is convex (concave),
then the function
\[
  f(\vec{x}) = \sum_{i=1}^K \theta(x_i)
\]
is Schur-convex (Schur-concave). \label{lemma:Schur}
\end{lemma}

A useful criterion for Schur-convexity is as follows.
\begin{lemma}[Schur's criterion]
Suppose that $F: \mathbb{R}_+^K \rightarrow \mathbb{R}$ is
differentiable and symmetric, meaning that $F(x_1,\ldots, x_K) =
F(x_{\pi(1)}, \ldots, x_{\pi(K)})$ for any $\vec{x}$ and permutation
$\pi$ of $\{1,\ldots, K\}$. Then
\begin{enumerate}
\item $F$ is Schur-convex if $(x_i - x_j) \left(  \frac{\partial F}{\partial x_i}(\vec{x}) -  \frac{\partial F}{\partial x_j}(\vec{x})
\right) \geq 0$ for all $i$ and $j$,
\item $F$ is Schur-concave if $(x_i - x_j) \left(  \frac{\partial F}{\partial x_i}(\vec{x}) -  \frac{\partial F}{\partial x_j}(\vec{x})
\right) \leq 0$ for all $i$ and $j$.
\end{enumerate}
\end{lemma}

See \cite{HLP52} for more on the theory of majorization and
Schur-convexity.

\section{Capacity-Fairness Tradeoff}
\label{results}

Usually there is a tradeoff between sum rate and fairness. The next
theorem illustrates such a tradeoff. It shows that the requirement
of symmetric fairness is more stringent than proportional fairness
in the sense that the symmetric capacity is always less than or
equal to the proportional-fair capacity. We use the notation $\Csum$
for the sum capacity, defined as
\[
\Csum(\mathcal{R}) = \max_{(r_1, \ldots, r_K) \in \mathcal{R}}
\sum_{i=1}^K r_i.
\]

\begin{theorem}
For any convex region $\mathcal{R}$,
\[\Csym(\mathcal{R}) \leq
\Cpf(\mathcal{R}) \leq \Csum(\mathcal{R}),
\]
\[\Csym(\mathcal{R}) \leq
\Cmm(\mathcal{R}) \leq \Csum(\mathcal{R}).
\]
Equality holds in the first inequality in the first line only if the
PF allocation is symmetric.
\end{theorem}

\begin{proof}
All inequalities are obvious except the the first inequality in the
first line, i.e., $\Csym(\mathcal{R}) \leq \Cpf(\mathcal{R})$.

 Let $(r_0,\ldots, r_0)$ be the maximal symmetric rate
allocation, and $(r_1^{PF},\ldots,r_K^{PF})$ be the proportional
fair rate allocation in region $\mathcal{R}$. We have
\[
  r_0^K \leq \prod_{i=1}^K r_i^{PF} \leq \Big( \frac{1}{K} \sum_{i=1}^K r_i^{PF} \Big)^K.
\]
The first inequality comes from the defining property of
proportional fairness, and the second is the AM-GM inequality. Thus,
\[
 r_0K \leq \sum_{i=1}^K r_i^{PF}.
\]
When equality holds, then we must have $r_1^{PF} = r_2^{PF} = \ldots
= r_K^{PF}$. This proves the first inequality.
\end{proof}

In general, the max-min capacity may or may not be larger than the
proportional fair capacity. However, in Gaussian MAC, both max-min
and proportional fair capacity achieve the sum capacity. In fact,
the max-min fair and proportional fair rate allocation in Gaussian
MAC coincide. It is hence not necessary to distinguish between
max-min and proportional fairness in Gaussian MAC. Meanwhile, in
Gaussian BC, max-min capacity is the same as symmetric capacity.

Given a rate allocation $\vec{r}$, we define its {\em efficiency},
$\eta$, as the ratio between the sum rate at $\vec{r}$ and the
maximal sum capacity. Obviously, this is a number between zero and
one. For proportional fairness, we have the following lower bound on
$\eta$.

\begin{theorem}
Let $\mathcal{R}$ be a convex region in $\mathbb{R}_+^K$, the
efficiency for proportional fairness, $\eta_{PF}$, is lower bounded
by~$1/K$.
\end{theorem}

\begin{proof}
By the definition of proportional fairness, the region $\mathcal{R}$
is contained in the polyhedron $\mathcal{P}$ defined by
\[
 \sum_{i=1}^K \frac{r_i}{r_i^{PF}} \leq K,
\]
and $r_i \geq 0 $ for all $i$. Therefore, $\Csum(\mathcal{R})$ is
not larger than $\Csum(\mathcal{P})$, which is equal to $K \max_i
r_i^{PF}$. Hence,
\[
\eta_{PF} = \frac{\Cpf(\mathcal{R})}{\Csum(\mathcal{R})} \geq
\frac{r_1^{PF} + \cdots + r_K^{PF}}{K \max_i r_i^{PF}} >
\frac{1}{K}.
\]
\end{proof}

We will show that in Gaussian MAC, $\eta_{PF}$ is exactly 1; it
achieves the maximal value. In Gaussian BC, the lower bound $1/K$ is
attained, i.e., $\inf \Cpf/\Csum = 1/K$ with the infimum taken over
all $K$-user Gaussian BC.

The major results are summarized in Table~\ref{table:MAC}
and~\ref{table:BC}. For both Gaussian MAC and BC, we also devise
efficient algorithms that compute the fair solutions.

\begin{table}
\begin{center}
\begin{tabular}{|c||c|c|c|}
\hline
 & Symmetric & Max-min and PF & Max. sum rate \\
 \hline\hline
$\eta$ & [0,1] & 1 & 1\\ \hline
 $C_*$ is Schur concave in $\vec{P}$
 &Theorem~\ref{thm:_MAC_sym_Schur}& trivial & trivial
\\
\hline
 $P_1 \leq \ldots \leq P_K \Rightarrow r^*_1\leq \ldots \leq r^*_K$ & trivial &
Theorem~\ref{theorem:order} & N/A \\ \hline

$\vec{P}\preceq \tilde{\vec{P}} \Rightarrow \vec{r}^* \preceq
\tilde{\vec{r}}^*$ & N/A & Corollary~\ref{cor:MAC_majorization} &
N/A \\ \hline
\end{tabular}
\end{center}
\caption{(Summary of results for Gaussian MAC) Capacities as a
function of the power constraint vector $\vec{P}$ are Schur concave.
For max-min fair or proportional fair, the rates are in increasing
order if the power constraints are sorted in increasing order. If
the power constraints are more spread out, so does the associated
rate allocation. } \label{table:MAC}
\end{table}

\begin{table}
\begin{center}
\begin{tabular}{|c||c|c|c|}
\hline
 & Symmetric and  Max-min & PF & Max. sum rate \\
 \hline\hline
$\eta$ & [0,1] & $[1/K, 1]$ & 1 \\
\hline $C_*$ is Schur convex in $\vec{N}$
 &Theorem~\ref{theorem:BC_sym_Schur}& Conjecture \ref{conjecture} & trivial
\\
\hline
 $N_1 \leq \ldots \leq N_K \Rightarrow r^*_1 \geq \ldots \geq r^*_K$ & trivial &
Corollary~\ref{cor:BC_PF_rate_order} & trivial \\ \hline

$N_1 \leq \ldots \leq N_K \Rightarrow p^*_1 \leq \ldots \leq p^*_K$
 & Theorem~\ref{theorem:BC_sym_power_order} & Theorem~\ref{theorem:BC_PF_power_order} & false \\
\hline
\end{tabular}
\end{center}
 \caption{(Summary of results for
Gaussian BC) Capacities as a function of the noise vector $\vec{N}$
are Schur convex. When the noise powers are sorted in increasing
order, then the data rates are in decreasing order, and the power
are in increasing order.  } \label{table:BC}
\end{table}

\section{Multiple-Access channels}
\label{sec:MAC} In a scalar Gaussian MAC with $K$ users, the
received signal is
\[
  Y = \sum_{i=1}^K X_i + Z,
\]
where $X_i$ is zero-mean Guassian with variance at most $P_i$ and
$Z$ is Gaussian noise with power $N$. Let $C$ denote the Shannon
capacity formula
\[
  C(x) = \frac{1}{2} \log (1+x).
\]
The capacity region of a scalar Gaussian MAC is \cite{CoverThomas}
\[
 \Big\{\vec{R}\in\mathbb{R}_+^K:\, \sum_{i \in S} R_i \leq C\Big(
   \frac{1}{N}\sum_{i\in S} P_i \Big),\  \text{for all } S\subseteq\Omega \Big\}
\]
The faces of the capacity region are hyper-planes in the form
$\sum_{i\in S} R_i = c$ for some subset $S$ of $\Omega$ and
constant~$c$.

In vector Gaussian MAC, the received signal is
\[
  \vec{Y} = \sum_{i=1}^K X_i \vec{s}_i + \vec{Z}.
\]
The $\vec{s}_i$'s are unit-norm column vectors of length~$L$, $X_i$
is a Gaussian random variable with zero mean and variance at most
$P_i$, and $\vec{Z}$ is the Gaussian noise vector with zero mean and
covariance matrix $N \mat{I}$, where $\mat{I}$ is the $L\times L$
identity matrix. The capacity region of a vector Gaussian MAC
is~\cite{VA99}
\[
 \Big\{ \vec{r}\in\mathbb{R}_+^K:\, \sum_{i\in S} r_i \leq \frac{1}{2}
 \log \Big| \mat{I} + \frac{1}{N} \sum_{i\in S} P_i s_i s_i^T \Big|\ \text{for all } S \subseteq
 \Omega \Big\}.
\]

\begin{figure}
\begin{center}
\includegraphics[width=70mm]{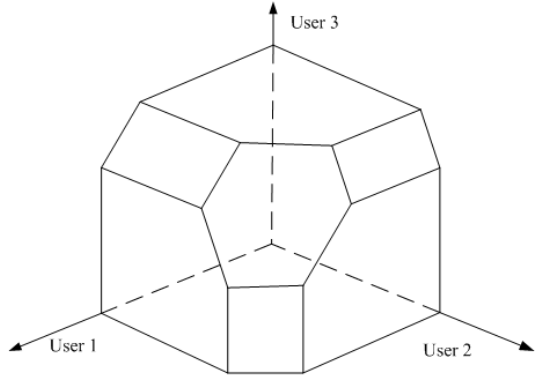}
\end{center}
\caption{Capacity Region of a 3-user Gaussian MAC channel}
\label{fig:capacity_region}
\end{figure}

A typical capacity region of Gaussian MAC for 3 users is illustrated
in Figure~\ref{fig:capacity_region}.

As in~\cite{TH98}, we will unify our treatment and present the
results for MAC in more general terms. For any subset
$S\subseteq\Omega$ and $\vec{r} = (r_1,\ldots, r_K)$, we use the
shorthand notation
\[
  \vec{r}(S) := \sum_{i\in S} r_i
\]
for the summation of components of $\vec{r}$ with indices in~$S$.

For any function $g$ mapping subsets of $\Omega$ to $\mathbb{R}_+$,
let $\mathcal{P}(g)$ denote the polyhedron
\[
 \Big\{ \vec{r}\in\mathbb{R}_+^K:\, r(S) \leq g(S),\ \forall S\subseteq \Omega
 \Big\}.
\]

If the function $g$ satisfies (i) $g(\emptyset) = 0$, (ii) $g(S)\leq
g(T)$ if $S\subseteq T$, and (iii) $g(S)+g(T)\geq g(S\cap T) +
g(S\cup T)$ for all subsets $S$ and $T$ of $\Omega$, then $g$ is
called a {\em rank function}. Property (ii) and (iii) are called the
{\em monotonic} and {\em submodular} property respectively. We say
that the polyhedron $\mathcal{P}(g)$ is a {\em polymatroid} when $g$
is a rank function.

The capacity regions of scalar and vector Gaussian MAC are
polymatroids. For scalar Gaussian MAC, we define the rank function
as
\begin{equation}
  g(S) = \frac{1}{2} \log\Big( 1+ \frac{1}{N} \sum_{i\in S} P_i
  \Big).
  \label{eq:MAC_rank_function}
\end{equation}

On the vector case, we define the rank function as
\begin{equation}
  g(S) = \frac{1}{2} \log \Big| \mat{I} + \frac{1}{N} \sum_{i\in S} P_i
  \vec{s}_i \vec{s}_i^T
  \Big|.
  \label{eq:MAC_rank_function2}
\end{equation}

The collection of points that achieve equality on total data rate
$\vec{r}(\Omega) = g(\Omega)$ is called the {\em dominant face}.
Given a vector $\vec{r}$ in $\mathcal{P}(g)$, we say that the set
$S$ is a {\em bottelneck} of $\vec{r}$ if
\[
  \vec{r}(S)= g(S).
\]
For any function $g:2^\Omega \rightarrow \mathbb{R}_+$, if
\[
g(A) \leq g(B) \text{ implies }g(A\cup C) \leq g(B\cup C)
\]
whenever $A\cap C = \emptyset = B\cap C$ and $|A| = |B|$, then we
say that $g$ satisfies the {\em order property}. The heuristic
meaning is as follow. If the achievable sum rate of group $A$ is
less than or equal to that of group $B$, then even if they cooperate
with users in group $C$, the sum rate of group $A\cup C$ is still
less than that of the group~$B \cup C$.

\paragraph{Example 1} Consider the function $g$ defined as $g(\{a\}) = 1$, $g(\{b\}) = 2$,
$g(\{c\})= g(\{b,c\}) = g(\{a,b\}) = 3$, $g(\{a,c\}) =
g(\{a,b,c\})=4$. We have $g(\{a\}) < g(\{b\})$ but $g(\{a,c\})>
g(\{b,c\})$. This function does not satisfy the order property.

\bigskip

The scalar Gaussian MAC satisfies the order property, but the vector
Gaussian MAC in general does not.

In both the scalar and vector Gaussian MAC, there is a canonical
choice of disagreement point for the Nash bargaining solution. Each
user can treat the signal of the others as noise and decode
independently. For scalar Gaussian MAC, the resulting data rate for
user $i$ is
\[
C \Big( \frac{P_i}{N+\sum_{j\neq i} P_j}  \Big).
\]
In the vector case, if user $i$ uses linear MMSE receiver with no
joint processing with others, the data rate is~\cite{VAT99}
\[
 C\Big( \frac{1}{N} P_i \vec{s}_i^T \mat{M}_i^{-1} \vec{s}_i
 \Big) ,
\]
where $\mat{M}_i$ is the matrix
\[
 \mat{I} + \frac{1}{N} \sum_{j\neq i} P_j \vec{s}_j^T \vec{s}_j .
\]

In both cases, they can be expressed in terms of the rank function
as
\begin{equation}
 d_i^* := g(\Omega) - g(\Omega\setminus\{i\}). \label{eq:canonical}
\end{equation}
The rate vector $\vec{d}^* = (d_1^*, \ldots, d_K^*)$ is called the
{\em canonical disagreement point}.

The next lemma is a useful consequence of the order property. The
proof is straightforward and is omitted.

\begin{lemma}
Let $g:2^\Omega \rightarrow \mathbb{R}_+$ be a function that
satisfies the order property.
\begin{enumerate}
\item If  $g(\{1\}) \leq g(\{2\}) \leq \ldots \leq g(\{K\})$ are in nondecreasing order,
then $g(\{1,2,\ldots, i\}) \leq g(S)$ for all $S\subseteq \Omega$ of
size $i$.

\item For any subset $A$, the function
$ g'(S) = g(S\cup A) - g(A)$ defined for $S\subseteq \Omega\setminus
A$ also satisfies the order property.
\end{enumerate}
\label{lemma:order}
\end{lemma}

\subsection{Symmetric Rate Allocation}

The computation of the symmetric capacity in $\mathcal{P}(g)$
amounts to finding the tightest constraint among $r(S)\leq g(S)$ for
all subsets $S$. Each component of the symmetric rate cannot exceed
$g(S)/|S|$ for all $S\subseteq \Omega$. The symmetric rate
allocation can be computed by checking $2^K-1$ constraints.

If $g$ satisfies the order property, the symmetric capacity can be
computed more efficiently using Lemma~\ref{lemma:order}. The
computation requires taking the minimum of only $K$ numbers.

\begin{theorem}
Let $g: 2^\Omega \rightarrow \mathbb{R}_+$ be a function that
satisfies the order property. By relabeling we can assume that
$g(\{i\}) \leq g(\{j\})$ whenever $i<j$. The symmetric capacity in
the polyhedron $\mathcal{P}(g)$ equals
\[
 K \cdot \min \Big\{ \frac{1}{k} g(\{1,\ldots, k\}) :\, k=1,\ldots, K
  \Big\}.
\]
\end{theorem}

\begin{corollary}
The infimum of fair efficiency of symmetric fairness, taken over all
$K$-user MAC, is zero.
\end{corollary}

\begin{proof}
Suppose that the power of user 1, $P_1$, is much less than the
others, so that
\[
 g(\{1\}) = \min\Big\{ \frac{1}{k} g(\{1,\ldots, k\}) :\, k=1,\ldots, K
  \Big\}.
\]
We see that the symmetric capacity approaches zero when $P_1$
approaches zero, while the maximal sum approaches a positive
constant.
\end{proof}

We next compare two MACs with different power constraints. If the
power constraints $P_1, \ldots, P_K$ become more disperse, then the
symmetric capacity will decrease.

\begin{theorem}
Let $\Csym(\vec{P},N)$ be the symmetric capacity of a scalar
Gaussian MAC with power constraints $\vec{P}$ and noise power~$N$.
If $\vec{P} \preceq \vec{P}'$, then $\Csym(\vec{P},N) \geq
\Csym(\vec{P}',N)$, i.e. the symmetric capacity of multiple-access
channel with fixed noise power is a Schur-concave function.
\label{thm:_MAC_sym_Schur}
\end{theorem}

\begin{proof}
Without loss of generality, we assume that the power constraints are
sorted in nondecreasing order, $P_1 \leq P_2 \leq \ldots \leq P_K$,
and $P_1' \leq P_2' \leq \ldots \leq P_K'$. Since scalar Gaussian
MAC satisfies the order property, the corresponding symmetric rate
allocations $\vec{r}^{sym}$ and $\vec{s}^{sym}$ are given by
\[
  r_i^{sym} = \min\Big\{ \frac{1}{k}  C\big(\frac{1}{N} \sum_{i=1}^k P_i\big):\, k=1,\ldots, K\Big\}
\]
and
\[
  s_i^{sym} = \min\Big\{ \frac{1}{k} C\big(\frac{1}{N}\sum_{i=1}^k P_i'\big):\, k=1,\ldots, K\Big\}
\]
for all $i$. As
\[
 \sum_{i=1}^k P_i \geq \sum_{i=1}^k P_i'
\]
for all $k$, we have
\[
 C\Big( \frac{1}{N} \sum_{i=1}^k P_i \Big) \geq C\Big( \frac{1}{N} \sum_{i=1}^k P_i' \Big).
\]
Therefore $r_i^{sym} \geq s_i^{sym}$.
\end{proof}

\subsection{Max-min and Proportional Fair Rate Allocation}

The following is a useful characterization of max-min
fairness~\cite{BG87}.

\begin{lemma}
For any function $g:2^\Omega \rightarrow \mathbb{R}_+$, a vector
$\vec{r}$ is max-min fair in the polyhedron $\mathcal{P}(g)$ if and
only if for all $i$, the $i$th component is largest in some
bottleneck. In other words, for $i=1,\ldots, K$, $i$ is contained in
a bottleneck $B$ and $r_i \geq r_j$ for all $j\in B$.
\label{lemma:maxmin}
\end{lemma}

The proof of Lemma~\ref{lemma:maxmin} is contained in the Appendix.
The next theorem is the main theorem in this section. The proof of
Theorem~\ref{theorem:MMPF} is in the Appendix (See
Prop.~\ref{prop:dominant_face} and~\ref{prop:nested}.)

\begin{theorem}
Suppose that $g: 2^\Omega \rightarrow \mathbb{R}_+$ satisfies the
submodular property. The max-min fair point $\vec{r}^{MM}$ in
$\mathcal{P}(g)$ is on the dominant face, and is majorized by every
point on the dominant face. \label{theorem:MMPF}
\end{theorem}

It is noted that in the theorem we do not assume that the function
$g$ is a rank function. The result holds as long as we have the
submodular property. The max-min fair allocation is fairer than any
other point on the dominant face in the sense of fairness induced by
majorization. The max-min fair solution also has the following
interpretations.

\begin{corollary}
If $g: 2^\Omega \rightarrow \mathcal{R}_+^K$ satisfies the
submodular property, the order property and $g(\emptyset)=0$, then
the max-min fair (and hence the proportional fair) solution
$\vec{r}^*$ maximizes $r_{[1]}$, $r_{[1]}+r_{[2]}, \ldots ,$ and
$\sum_{j=1}^{K-1} r_{[j]}$ simultaneously over all points on the
dominant face.
\end{corollary}

\begin{corollary}
If $g$ satisfies the submodular property, then the max-min point and
the proportional fair point in $\mathcal{P}(g)$ coincide.
\label{cor:maxmin_PF}
\end{corollary}

\begin{proof}
Since $\log$ is a concave function, $\sum_i \log(r_i)$ is
Schur-concave by Lemma~\ref{lemma:Schur}. Because the max-min point
$\vec{r}^{MM}$ is majorized by any point $\vec{r}$ on the dominant
face, we have
\[
\sum_i \log(r_i^{MM}) \geq \sum_i \log(r_i).
\]
Hence $\vec{r}^{PF} = \vec{r}^{MM}$.
\end{proof}

\begin{corollary}
If $g$ satisfies the submodular property, then the max-min (and the
proportional fair) point in $\mathcal{P}(g)$ is the point on the
dominant face that minimizes the Euclidean norm.
\end{corollary}

\begin{proof}
The function $f(x) = x^2$ is convex. The proof is similar to the
proof of the last corollary.
\end{proof}

\begin{corollary}
If $g$ satisfies the submodular property, the fairness efficiency of
max-min fairness and proportional fairness is equal to 1.
\end{corollary}

\begin{theorem}
Suppose that $g:2^\Omega \rightarrow \mathbb{R}_+$ satisfies both
submodular and order property, and assume $g(\{1\}) \leq g(\{2\})
\leq \ldots \leq g(\{K\})$ after suitable relabeling. Then the
components of max-min fair (and hence the proportional fair)
solution $\vec{r}^*$ in $\mathcal{P}(g)$ are in nondecreasing order,
$r^*_1 \leq r^*_2 \leq \ldots \leq r_K^*$. \label{theorem:order}
\end{theorem}

\begin{proof}
Consider any $i$ in $\Omega$. The index $i$ is contained in a
bottleneck $A_i$ of $\vec{r}^*$ so that $r^*_i = \max \{ r^*_j:\,
j\in A_i\}$. If $i-1 \in A_i$, then $r^*_{i-1} \leq r^*_i$.
Otherwise, suppose $i-1 \not\in A_i$ and let $S$ denote $A_i
\setminus \{i\}$. Then
\[ \vec{r}^*(S \cup \{i-1\}) \leq g(S \cup \{i -1 \}) \leq  g(S \cup \{i\})
  =\vec{r}^*(S \cup \{i\}).
\]
We have used the order property in the second inequality. This
implies that $ r^*_{i-1} \leq r^*_i$.
\end{proof}

A typical class of functions that satisfies the order property is
the generalized symmetric functions. A rank function $g$ is said to
be {\em generalized symmetric} if it has the form
\begin{equation}
  g(S) = \phi( \vec{Q}(S) ), \label{eq:generalized_sym}
\end{equation}
where $\phi$ is a monotonic increasing and concave function with
$\phi(0)=0$ and $\vec{Q} \in \mathbb{R}_+^K$. The rank function in
the scalar MAC is an example of generalized symmetric function.

The next theorem compares two Gaussian MACs, with the same total
power but different distribution in the power constraints. It shows
that if the distribution of power constraints is more spread out, so
does the corresponding max-min fair rate allocation. The proof is
relegated to the appendix.

\begin{theorem}
Let $g$ be a generalized symmetric rank function defined as $g(S) =
\phi(\vec{P}(S))$, for some vector $\vec{P} \in \mathbb{R}_+^K$. Let
$\tilde{\vec{P}}$ be a vector that majorizes $\vec{P}$, and $g'$ be
the generalized symmetric rank function $g'(S) =
\phi(\tilde{\vec{P}}(S))$. Then the max-min fair capacity associated
to $\vec{P}$ is the same as the max-min fair capacity associated
with $\tilde{\vec{P}}$, and he max-min fair point in
$\mathcal{P}(g)$ is majorized by the max-min fair point in
$\mathcal{P}(g')$. \label{theorem:MAC_majorization}
\end{theorem}

\begin{corollary}
Let $\vec{r}^*(\vec{P},n)$ be max-min fair in a Gaussian MAC with
power constraints $P_1 \leq \ldots \leq P_K$ and noise power $n$. If
$\vec{P} \preceq \vec{P}'$, then $\vec{r}^*(\vec{P},n) \preceq
\vec{r}^*(\vec{P}',n)$. \label{cor:MAC_majorization}
\end{corollary}

\subsection{Algorithm}

We present a general recursive algorithm that computes the fair
solution in Gaussian MAC. It is a variation of the algorithm
in~\cite[p.527]{BG87}, which computes the max-min fair rate vector
in flow control problem. The algorithm to be described below
exploits the submodular property, and has shorter running time. The
basic idea is contained in the next proposition.

\begin{proposition}
Let $g$ be a function mapping $2^\Omega$ to $\mathbb{R}_+$, and let
$S_0$ be a subset of $\Omega$ that achieves the minimum
\[
 \min_{\emptyset \neq S\subseteq \Omega} g(S)/|S|.
\]
Let $\vec{r}^*$ denote the the max-min or proportional fair point
(they are the same by Corollary~\ref{cor:maxmin_PF})
in~$\mathcal{P}(g)$, we have $r^*_i \geq g(S_0)/|S_0|$ for all $i$,
with equality when $i \in S_0$. \label{prop:slow_algorithm}
\end{proposition}

\begin{proof}
For each $i$, $i$ is contained in a bottleneck $B_i$ of $\vec{r}^*$,
so that $r_i \geq r_j$ for all $j\in B_i$. So $r_i$ must be larger
than or equal to the average $\vec{r}^*(B_i) /|B_i|$, and thereby
\[
 r_i^*  \geq \frac{1}{|B_i|}\sum_{j\in B_i} r_j = g(B_i)/|B_i| \geq g(S_0)/|S_0|.
\]
Therefore $r^*_i \geq g(S_0)/|S_0|$ for all $i$. Summing over all
$i\in S_0$, we obtain
\[
 \vec{r}^*(S_0) \geq g(S_0).
\]
We must have equality in all the above inequalities. In particular,
$r^*_i = g(S_0)/|S_0|$ for all $i\in S_0$. The set $S_0$ is in fact
a bottleneck of $\vec{r}^*$.
\end{proof}

This motivates the max-min algorithm.

\paragraph{Max-min algorithm} The algorithm starts by first obtaining the subset $S_0\in\Omega$
described in Proposition~\ref{prop:slow_algorithm}, and set the rate
of users in $S_0$ to $g(S_0)/|S_0|$. The rate of other users are
computed by recursively applying the above computation to $\Omega'
:= \Omega\setminus S_0$ with
\begin{equation}
 g'(S) := g(S \cup S_0) - g(S_0) \label{eq:recursion}
\end{equation}
for $S \subseteq \Omega'$.

{\em Remark:} It is noted that we only need the submodular property
in proving the correctness of the max-min algorithm. The function
$g$ need not satisfy the monotonic property or $g(\emptyset)=0$. We
will use the following lemma in proving the correctness of the
algorithm. Note that the lemma holds in general for arbitrary $g$.

\begin{lemma}
Let $g$ be a function from $2^\Omega$ to $\mathbb{R}_+$, and $S_0
\subseteq \Omega$ be chosen such that
\[g(S_0)/|S_0| = \min_{
\emptyset \neq S \subseteq \Omega} g(S)/|S|.\]
Define the function
\[
 g'(S) := g(S\cup S_0) - g(S_0)
\]
for $S \subseteq \Omega' := \Omega \setminus S_0$.

\begin{enumerate}
\item (Non-negativity) $g' \geq 0$, for all $S \subseteq \Omega'$. \label{item1}

\item (Extension of bottleneck) Let $\vec{r}$ be a vector in $\mathcal{P}(g)$ such that $\vec{r}(S_0) = g(S_0)$.
 Let $\vec{r}'$ be the restriction of the vector
$\vec{r}$ on $\Omega'$. If $B'$ is a bottleneck of $\vec{r}'$ in
$\mathcal{P}(g')$, then $B' \cup S_0$ is a bottleneck of $\vec{r}$
in $\mathcal{P}(g)$. \label{item2}

\item (Preservation of order property) If $g$ satisfies the order property, so does~$g'$.
\label{item3}
\end{enumerate}
\end{lemma}

\begin{proof}
(\ref{item1})  By construction, we have
\[
 g(S \cup S_0) \geq g(S_0) \frac{|S\cup S_0|}{|S_0|}.
 \]
Hence $g(S \cup S_0) \geq g(S_0)$.

(\ref{item2}) $\vec{r}'(B')$ equals $g'(B')$, which is $g(B' \cup
S_0) - g(S_0)$ by definition. Therefore
\begin{align*}
g(B'\cup S_0) &= \vec{r}'(B') + g(S_0) \\
&= \vec{r}(B') + \vec{r}(S_0) \\
&= \vec{r}(B' \cup S_0),
\end{align*}
and $B' \cup A$ is a bottleneck of $\vec{r}$ in $\mathcal{P}(g)$.

(\ref{item3})  Let $S$, $T$ and $U$ be subsets of $\Omega\setminus
S_0$ so that $|S|=|T|$ and $S\cap U = \emptyset = T\cap U$. If
$g'(S) \leq g'(T)$, then
\begin{align*}
 g'(S \cup U) &= g(S \cup U \cup A) - g(A) \\
 & \leq g(T \cup U \cup A) - g(A) \\
 &= g'(T \cup U).
\end{align*}
\end{proof}

\begin{theorem}
Suppose $g: 2^\Omega \rightarrow \mathbb{R}_+$ satisfies the
submodular property. In the polyhedron $\mathcal{P}(g)$, the result
obtained by the max-min algorithm is the max-min fair vector.
\end{theorem}

\begin{proof}
Let $\vec{r}$ be the vector returned by the max-min algorithm, and
let $S_0$ be a subset of $\Omega$ such that $g(S_0)/|S_0| =
\min_{\emptyset \neq S \subseteq \Omega} g(S)/|S|$. If $S_0 =
\Omega$, then the components in $\vec{r}$ are constant and equal
$g(S_0)/|S_0|$. It belongs to $\mathcal{P}(g)$ because for any
$S\subseteq \Omega$
\begin{equation}
  \vec{r}(S) = |S| \frac{g(S_0)}{|S_0|} \leq g(S).
  \label{eq:belongs_to_P}
\end{equation}
It is easy to see that $\vec{r}$ is max-min fair.

Otherwise, if $S_0 \subsetneq \Omega$, then we have to apply the
algorithm recursively.  In this case, the vector $\vec{r}$ satisfies
the following properties: (i) $r_i = g(S_0)/|S_0|$ for all $i\in
S_0$ and (ii) the components of $\vec{r}$ with index in
$\Omega\setminus S_0$ yield the max-min solution to the polymatroid
$\mathcal{M}'$ on $\Omega\setminus S_0$ with rank function $g'$
defined as in~\eqref{eq:recursion}. It is easy to check that $g'$
also satisfies the submodular property. We first verify that
$\vec{r}$ is in $\mathcal{P}(g)$. For any subset $A\subseteq
\Omega$, we can write $A$ as $A_1 \cup A_2$ with $A_1 \subseteq S_0$
and $A_2 \cap S_0 = \emptyset$. We decompose $\vec{r}(A)$ as
\begin{align*}
\vec{r}(A) &= \vec{r}(A_1) + \vec{r}(A_2) \\
&\leq |A_1| \frac{g(S_0)}{|S_0|} + g(A_2 \cup S_0) - g(S_0) \\
& \leq g(A_1) + g(A_2 \cup S_0) - g(S_0)
\end{align*}
In the last inequality, we have used  the defining property of
$S_0$, i.e., $|A_1| g(S_0)/|S_0| \leq g(A_1)$. By the submodularity
of $g$, we have
\[
 g(S_0) + g(A_1 \cup A_2) \geq g(A_1) + g(A_2 \cup S_0).
\]
Therefore $\vec{r}(A) \leq g(A_1 \cup A_2)$, and thus $\vec{r}$ is
in $\mathcal{P}(g)$.

We now show that for each $i=1,\ldots, K$, $i$ is in some bottleneck
$A_i$ such that $r_i = \max\{r_j:\, j\in A_i \}$. We will apply
Lemma~\ref{lemma:maxmin} and conclude that $\vec{r}$ is the max-min
vector in~$\mathcal{P}(g)$. For $i\in S_0$, we can take $S_0$ as the
required bottleneck $A_i$. For $i\not\in S_0$, $i$ is an element of
some bottleneck $B'$ in the polyhedron $\mathcal{P}(g')$ such that
$r_i \geq r_j$ for all $j\in B'$. By part~(\ref{item2}) in the
previous lemma, $B'\cup S_0$ is a bottleneck of~$\mathcal{P}(g')$.
By Prop.~\ref{prop:slow_algorithm}, we can show that $r_i \geq r_j$
for all $j \in S_0$. Indeed,
\begin{align*}
 r_i & \geq \frac{g'(B')}{|B'|} \\
 &=\frac{1}{|B'|}( g(B' \cup S_0) -
 g(S_0)) \\
 & \geq \frac{1}{|B'|} \left( g(S_0) \frac{|S_0 \cup B'|}{|S_0|} - g(S_0) \right) \\
 & = \frac{g(S_0)}{|S_0|} = r_j.
\end{align*}

Therefore $i$ is in the bottleneck $S_0\cup B'$ and $r_i \geq r_j$
for all $j\in B'\cup S_0$. The vector $\vec{r}$ is thereby max-min
fair by Lemma~\ref{lemma:maxmin}
\end{proof}

\paragraph{Example 1 (continued)} We compute the max-min fair vector
in Example 1. The minimum
\[
  \min_{\emptyset \neq S \subseteq \Omega} g(S)/|S|
\]
is achieved when $S = \{a\}$. We set $r_1^{MM} = g(\{a\}) = 1$. Next
define
\[
  g'(S) := g(\{a\} \cup S) - g(\{a\})
\]
for $S \subseteq \{b,c\}$. Now $g'(\{b\}) = 2$ and $g'(\{c\}) = 3 =
g'(\{b,c\})$. So in this recursive step, we have the minimum
\[
 g'(\{b,c\})/2 = 3/2.
\]
The resulting max-min fair solution is
\[
 \vec{r}^{MM} = (1, 3/2, 3/2).
\]

\bigskip

In the max-min algorithm, if we compute the minimum of $g(S)/|S|$,
$\emptyset \neq S \subseteq \Omega$, in a straightforward manner by
comparing $2^K-1$ numbers, the complexity of the algorithm is
exponential in the number of users. A more efficient implementation
was described in~\cite{MMK06} if $g$ is a rank function. However,
when the function $g$ satisfies the order property, we have a much
faster algorithm.

\begin{proposition}
If the function $g$ satisfies the order property and the submodular
property, then the max-min fair point in $\mathcal{P}(g)$ can be
computed in $O(K^2)$ time.
\end{proposition}

\begin{proof}
In the max-min algorithm, instead of finding the minimum
\[
  \min_{\emptyset\neq S \subseteq \Omega} g(S)/|S|
\]
over all subsets of $\Omega$, we sort $g(\{1\}), g(\{2\}), \ldots,
g(\{K\})$ in nondecreasing order. This can be done in $O(K \log(K))$
time. For notational convenience, we relabel the users so that
\[
 g(\{1\}) \leq g(\{2\}) \leq \ldots \leq g(\{K\}).
\]

Since function $g$ satisfies the order property, for any $k$, the
minimum
\[
 \min \{ g(S)/k:\, S \subseteq \Omega, |S|=k \}
\]
is achieved by $\{1,2,\ldots, k\}$ by Lemma~\ref{lemma:order}.
Instead of comparing $g(S)/|S|$ over all subsets $S$ of $\Omega$, it
is sufficient to examine $g(\{1,\ldots, k\})/k$, for $k = 1,\ldots,
K$. The minimum can be found in $O(K)$ time. In the next recursion,
the function $g'(S) = g(S\cup S_0) - g(S_0)$ also satisfies the
order property, and
\[
 g'(\{k_0+1\}) \leq g'(\{k_0+2\}) \leq \ldots \leq g'(\{K\}).
\]
The recursion can continue without any further sorting. There are at
most $K$ recursive steps and each step takes $O(K)$ time. The total
complexity is therefore $O(K^2)$.
\end{proof}

\paragraph{Example 2} Consider a scalar Gaussian MAC with 4 users.
Their power constraints are 2, 8, 200, and 300. We let $\vec{P}$ be
the vector $(2, 8, 200, 300)$. The noise power at the receiver is
equal to 1. Let
\[
  g_1(S) := 0.5 \log (1+ \vec{P}(S) )
\]
for $S\subseteq \Omega$. Here we use the natural logarithm function.
We want to find the max-min fair rate allocation or the proportional
fair rate allocation in $\mathcal{P}(g_1)$. The function $g_1$ is a
rank function and satisfies the order property.

We first compute the minimum of $g_1(\{1\}$, $g_1(\{1,2\})/2$,
$g_1(\{1,2,3\})/3$ and $g_1(\{1,2,3,4\})/4$. The minimum is
$g_1(\{1\}) = 0.5493$. We set $r^{MM}_1 = 0.5493$.

In the next recursive step, set
\[ g_2(S) := g_1(S \cup \{1\}) - g_1(\{1\})
\]
for $S\subseteq \{2,3,4\}$. The minimum of $g_2(\{2\})$,
$g_2(\{2,3\})/2$ and $g_3(\{2,3,4\})/3$ is $g_2(\{2\})$. We set
$r^{MM}_2 = g_2(\{2\}) = 0.6496$.

Let
\[
 g_3(S) := g_2(S \cup \{2\})-g_2(\{2\}) = g_1(S \cup \{1,2\}) -
 g_1(\{1,2\})
\]
for $S \subseteq \{3,4\}$, and compute the minimum of $g_3(\{3\})$
and $g_3(\{3,4\})/2$. The minimum is $g_3(\{3,4\})2 = 0.9596$, and
we assign $0.9596$ to both $r^{MM}_3$ and $r^{MM}_4$.

The max-min fair rate allocation is thus
\[
 \vec{r}^{MM} = (0.5493, 0.6496, 0.9596, 0.9596).
\]

\bigskip

The computation of the Nash bargaining solution in $\mathcal{P}(g)$
with disagreement point $\vec{d}$ amounts to finding the max-min
fair solution in~$\mathcal{P}(g')$, where
\[
 g'(S) = g(S) - \vec{d}(S).
\]
It is noted that if $g$ is a rank function, the translated $g'$ in
general does not satisfy the monotonic property. However, the
max-min algorithm works without assuming the monotonic property. We
can apply the max-min algorithm to find the Nash bargaining solution
for any disagreement point.

We conclude this section by presenting an algorithm for computing
Nash bargaining solution when the rank function is generalized
symmetric.

\begin{lemma}
Let $g$ be a generalized symmetric rank function on $\Omega$,
\[
 g(S) = \phi( \vec{P}(S)),
\]
for $S\subseteq \Omega$, and $\vec{P}$ is a vector in
$\mathbb{R}_+^K$ such that.
\[
 P_1 \leq P_2 \leq \ldots \leq P_K.
\]
Then for any $k\in \{1,\ldots, K\}$, the minimum
\[
 \min \Big\{ g(S) + \sum_{i\in S} g(\Omega\setminus \{i\}) :\, S\subseteq \Omega, |S|=k \Big\}
\]
is equal to
\[
 g(\{1,2,\ldots, k\}) + \sum_{i=1}^k g(\Omega\setminus \{i\}).
\]
\end{lemma}

\begin{proof}
Let $\bar{P}_i$ denote $\sum_{j\neq i}^K P_j$, where the summation
is over all indices except~$i$. The lemma claims that
$\phi(\vec{P}(\{1,\ldots, k\} ) + \sum_{i=1}^k \phi(\bar{P}_i)$, is
the minimum. By Lemma~\ref{lemma:Schur}, it suffices to show that
\[ (\bar{P}_1, \bar{P}_2, \ldots, \bar{P}_k,
P_1 + \ldots+ P_k )
\]
is majorized by
\[ (\bar{P}_{i_1}, \bar{P}_{i_2}, \ldots, \bar{P}_{i_k},
P_{i_1} + \ldots+ P_{i_k} )
\]
for any choice of $i_1, \ldots, i_k$, $1 \leq i_1 < \ldots < i_k
\leq K$.

By subtracting $\sum_{i=1}^K P_i$ from both vectors, we only need to
show that
\[
 (P_1, P_2 ,\ldots, P_k, P_{k+1}+\ldots +P_K) \succeq
 \Big( P_{i_1}, P_{i_1} ,\ldots, P_{i_k}, \sum_{j\not\in B} P_j \Big).
\]
where $B$ denote the set $\{i_1, \ldots, i_k\}$.

Let $\vec{Q}$ be the vector $ \Big( P_{i_1}, P_{i_1} ,\ldots,
P_{i_k}, \sum_{j\not\in B} P_j \Big)$, and $Q_{[1]}$, $Q_{[2]},
\ldots, Q_{[K]}$ be the components of $\vec{Q}$ in nondecreasing
order. It is easy to see that
\[
 \sum_{i=1}^k P_i \leq Q_{[k]}.
\]
Therefore $(P_1, P_2 ,\ldots, P_k, P_{k+1}+\ldots +P_K)$ majorizes
$\vec{Q}$. This finishes the proof of the lemma.
\end{proof}

\begin{proposition}
For a generalized symmetric rank function $g$, the Nash bargaining
solution in $\mathcal{P}(g)$ with the canonical disagreement point
can be computed in $O(K^2)$ time.
\end{proposition}

\begin{proof}
Assume without loss of generality that the power constraints are
arranged in nondecreasing order. Let $\vec{d}^*$ denote the
canonical disagreement point, i.e., for $i=1,\ldots, K$, $ d_i^* :=
g(\Omega) - g(\Omega \setminus \{i\})$. Let
\[ h(S) := g(S) - \vec{d}^*(S) = g(S) + \sum_{i\in S} g(\Omega\setminus\{i\}) - |S|g(\Omega).
\]
We relabel the users so that $h(\{1\}) \leq \ldots \leq h(\{K \})$.
By the previous lemma, the minimum
\[
\min \Big\{ h(S) :\, S\subseteq \Omega, |S|=k \Big\}
\]
is $h(\{1,\ldots, k\})$. Therefore,
\[
 \min_{\emptyset \neq S \subseteq \Omega} h(S)/|S| =
 \min_{k=1,\ldots, K}  h(\{1,\ldots, k\})/k.
\]
The minimum can be obtained efficiently after sorting $h(\{1\}),
\ldots, h(\{K\})$. Suppose that the minimum is $h(\{1,\ldots,
i_0\})/|i_0|$. We set $r^{MM}_i$ to $h(\{1,\ldots, i_0\})/|i_0| +
d^*_i$ for $i=1,\ldots, i_0$.

We next show that the same procedure can be repeated in the next
recursive step. Let $S_0 = \{1,\ldots, i_0\}$, and $g'(S) = g(S\cup
S_0)-g(S_0)$ for $S\subseteq \Omega' := \{i_0+1, \ldots, K\}$. It is
easy to see that $g'$ is generalized symmetric. Also, we can verify
that $(d^*_{i_0+1}, \ldots, d^*_K)$ is the canonical disagreement
point for $g'$. Indeed,
\begin{align*}
 d^*_i &= g(\Omega) - g(\Omega \setminus \{i\}) \\
 &= g'(\Omega') - g'(\Omega' \setminus \{i\})
\end{align*}
for all $i\not\in S_0$.

Each recursive step takes $O(K))$. As there are at most $K$ steps,
the complexity for computing the Nash bargaining solution with the
canonical disagreement point is $O(K^2)$.
\end{proof}

\paragraph{Example 2 (cont'd)} We compute the Nash bargaining
solution in the MAC as in Example 2, with the canonical disagreement
point $\vec{d}^*$.
\begin{align*}
d^*_1 &= g_1(\{1,2,3,4\}) - g_1(\{2,3,4\}) = 0.0020 \\
d^*_2 &= g_1(\{1,2,3,4\}) - g_1(\{1,3,4\}) = 0.0079 \\
d^*_3 &= g_1(\{1,2,3,4\}) - g_1(\{1,2,4\}) = 0.2483 \\
d^*_4 &= g_1(\{1,2,3,4\}) - g_1(\{1,2,3\}) = 0.4423
\end{align*}

The minimum of
\[
 \big( g_1(\{1,\ldots, k\}) - d^*_1 - \ldots - d^*_k \big) / k
\]
for $k=1,2,3,4$, is $g_1(\{1\}) - d^*_1$. Therefore,
\[ r^{MM}_1 = (g_1(\{1\}) - d^*_1)+d^*_1 = 0.5493.
\]

For $S \subseteq \{2,3,4\}$, let $g_2(S) := g_1(S \cup \{1\}) -
g_1(\{1\})$.
\begin{align*}
 g_2(\{2\}) - d^*_2 &= 0.6418\\
 \frac{g_2(\{2,3\}) - d^*_2 - d^*_3}{2} &= 0.9352\\
 \frac{g_2(\{2,3,4\}) - d^*_2 - d^*_3- d^*_4}{3} &=  0.6235
\end{align*}
The last equation yields the bottleneck. We set
\begin{align*}
r^{MM}_2 &= 0.6235 + d^*_2
=  0.6314\\
r^{MM}_3 &= 0.6235 + d^*_3 = 0.8718 \\
 r^{MM}_4 &= 0.6235 + d^*_4 = 1.0657
\end{align*}

The resulting Nash bargaining solution is
\[
 (0.5493, 0.6314, 0.8718, 1.0657).
\]

\section{Broadcast Channels}
\label{sec:BC} In a $K$-user Gaussian broadcast channel, the
received signal of the $i$th user is
\[
  Y_i = X + Z_i
\]
where $X$ is a zero-mean Gaussian random variable with variance
$P_T$ and $Z_i$ is the noise at the $i$th receiver, which is modeled
as a Gaussian variable with mean zero and variance~$N_i$. We will
assume that $N_1 \leq N_2 \leq \ldots \leq N_K$ throughout this
section. Every point $\vec{r}$ on the boundary of the capacity
region satisfies
\[
 r_i = \frac{1}{2} \log\Big(1+ \frac{\alpha_i P_T}{N_i + \sum_{j=1}^{i-1}
  \alpha_j P_T}   \Big)
\]
for some $\alpha_1, \ldots, \alpha_K$ such that
\[
\sum_{j=1}^K\alpha_j=1, \ \alpha_j\geq 0.
\]

\subsection{Symmetric Capacity}

In order to obtain the symmetric capacity in a BC, we solve a
related problem of finding the power distribution so that the users
have a common SINR $\gamma$, i.e.,
\[
\gamma = \frac{p_1}{N_1} = \frac{p_2}{N_2 + p_1} = \cdots =
\frac{p_K}{N_K + p_{K-1} + \ldots + p_1}.
\]
The noise vector $\vec{N}$ can be expressed in terms of the power
vector $\vec{p}$ by a matrix multiplication
\[
\begin{bmatrix}N_1 \\ N_2 \\ \vdots \\ N_K \end{bmatrix} =
\begin{bmatrix}
1/\gamma    & 0 & \cdots& 0\\
-1  & 1/\gamma  & \cdots & 0 \\
\vdots & \vdots & \ddots & \vdots\\
-1 & -1 & -1 & 1/\gamma
\end{bmatrix}
\begin{bmatrix} p_1 \\ p_2 \\ \vdots \\ p_K \end{bmatrix}.
\]
Let $G_\gamma$ denote the lower triangular matrix in the above
equation. The diagonal elements of $G_\gamma$ all equal $1/\gamma$,
and the elements below the diagonal are all~$-1$. The following
lemma is obtained by straightforward calculation.

\begin{lemma}
The inverse of $G_\gamma$ is a non-negative matrix. The
$(i,j)$-entry of $G_\gamma^{-1}$ is
\[
[G_{\gamma}^{-1}]_{ij} = \begin{cases}
0 & \text{if }i< j, \\
\gamma & \text{if } i=j, \\
 \gamma^2 (1+\gamma)^{i-j-1}    & \text{otherwise}.
\end{cases}
\]
\end{lemma}

Hence, given the noise powers $\vec{N} = (N_1, \ldots, N_K)$ and the
SINR requirement $\gamma$, we get the corresponding power allocation
by multiplying $G_\gamma^{-1}$ by $\vec{N}$. The following theorem
is an immediate consequence. It says that in a Gaussian BC, users
with lower noise power uses less power.

\begin{theorem}
In a Gaussian BC with noise powers $N_1 \leq \ldots \leq N_K$, the
powers corresponding to the symmetric rate allocation are in
increasing order,
\[
 p_1^{sym} \leq p_2^{sym} \leq \ldots \leq p_K^{sym}.
\]
\label{theorem:BC_sym_power_order}
\end{theorem}

\begin{proof}
Let $\vec{p}$ be the power vector so that all users have SINR
$\gamma$, and let $h_{ij}$ be the $(i,j)$-entry of $G_\gamma^{-1}$.
The power of user $i$ can be obtained by
\[
  p_i = h_{i1}N_1 + \ldots + h_{ii}N_i.
\]
Since the first term is nonnegative, we can remove the first term
and get
\[
  p_i \geq h_{i2}N_2 + h_{i3}N_3 +\ldots + h_{ii}N_i.
\]
It is clear from the previous lemma that $h_{ij}$ depends only on
$i-j$, hence
\[
  p_i \geq h_{(i-1)1}N_2 + h_{(i-1)2} N_3 + \ldots + h_{(i-1)(i-1)}N_i
\]
We now use the assumption $N_1 \leq \ldots \leq N_K$ to obtain
\begin{align*}
  p_i &\geq h_{(i-1)1}N_1 + h_{(i-1)2} N_2 + \ldots + h_{(i-1)(i-1)}N_{i-1} \\
  &= p_{i-1}.
\end{align*}
Consequently $p_1 \leq p_2 \leq \ldots \leq p_K$.
\end{proof}

We denote the required total power by $\phi(\vec{N},\gamma)$, which
can be computed by
\begin{align*}
 \phi(\vec{N},\gamma) &= [1,\ldots, 1] \cdot G_\gamma^{-1} \vec{N}
 \\
 &= N_1\theta_1(\gamma) + N_2\theta_2(\gamma) + \ldots + N_K
 \theta_K(\gamma)
\end{align*}
where $\theta_j(\gamma)$ denote the sum of elements in the $j$th
column of $G_\gamma^{-1}$. It is noted that the function
$\theta_j(\gamma)$ is a convex function of $\gamma$ for all $j$.
Furthermore, we have
\begin{equation}
 \theta_1(\gamma) \geq \theta_2(\gamma) \geq \ldots \geq
 \theta_K(\gamma).
 \label{eq:decreasing_theta}
\end{equation}
Given a noise vector $\vec{N}$, the function $\phi$ is a convex and
monotonically increasing function of $\gamma$.

The next theorem compares the symmetric capacity of two broadcast
channels with the same total power constraint.

\begin{theorem}
If  $\vec{N} \preceq \vec{N}'$, then $\Csym(\vec{N},P_T) \leq
\Csym(\vec{N}',P_T)$. In other words, $\Csym(\cdot,P_T)$ is a
Schur-convex function. \label{theorem:BC_sym_Schur}
\end{theorem}

\begin{figure}
\begin{center}
\includegraphics[width=110mm]{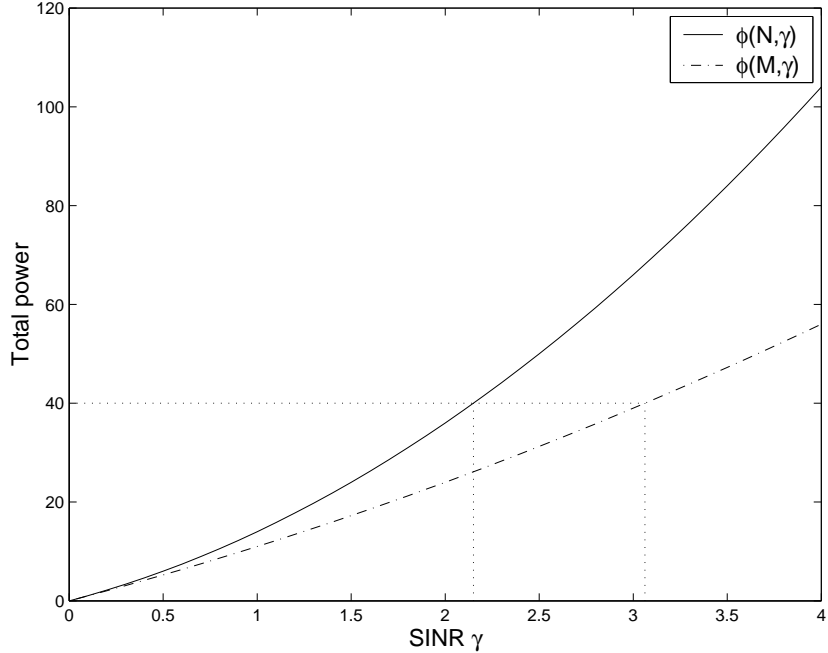}
\end{center}
\caption{The total power required for symmetric rate allocation as a
function of SINR. The noise vector associated to the lower curve
majorizes the noise vector associated to the upper one. The dash
lines show how to find the SINR when the total power is given.}
\label{fig:phi}
\end{figure}

\begin{proof}
Assume without loss of generality that the components of $\vec{N}_1$
and $\vec{N}_2$ are sorted in nondecreasing order. For a given total
power constraint and noise vector $\vec{N}$, we can obtain the SINR
by solving the equation
\[\phi(\vec{N},\gamma) = P_T.
\]
The proof is complete if we can show that $\phi(\vec{N},\gamma) \geq
\phi(\vec{N}',\gamma)$ for all $\gamma$, i.e., $\phi(\cdot,\gamma)$
is Schur-concave for all~$\gamma$ (Fig.~\ref{fig:phi}). Indeed, for
$i > j$, we have $N_i \geq N_j$ and
\[
  \frac{\partial \phi}{\partial N_i} - \frac{\partial \phi}{\partial
  N_j} = \theta_i(\gamma) - \theta_j(\gamma) \leq 0.
\]
by \eqref{eq:decreasing_theta}. Therefore $\phi(\cdot,\gamma)$ is
Schur-concave by Schur's criterion.
\end{proof}

\begin{theorem}
For fixed noise powers, the symmetric capacity is a convex function
of the total power.
\end{theorem}

\begin{proof}
For a fixed noise vector $\vec{N}$, the function
$\phi(\vec{N},\gamma)$ is a convex function of $\gamma$. Hence the
inverse function $\phi^{-1}$ is a concave function.
\end{proof}

\paragraph{Algorithm} We have a numerical algorithm computing the symmetric capacity in
Gaussian BC by means of the function $\phi(\vec{N},\gamma)$. For a
given total power $P_T$, we search for the value of $\gamma$ so that
$\phi(\vec{N},\gamma) = P_T$. This can be done easily as $\phi$ is a
monotonic function of $\gamma$. We then compute the data rate from
$\gamma$.

\begin{corollary}
The infimum of $\eta_{sym}$, taken over all $K$-user Gaussian BC, is
zero.
\end{corollary}

\begin{proof}
The sum capacity is attained if we allocate all power to user 1,
\[
 \Csum = \frac{1}{2} \log \Big( 1+ \frac{P_T}{N_1} \Big).
\]
Suppose the noise power of user $K$, $N_K$, is increased, while the
others are fixed, the value of the function $\phi(\vec{N},\gamma)$
is increased for all~$\gamma$. Then the symmetric capacity is
decreased, but the sum capacity remains constant. By taking $N_K$
approaching infinity, $\eta_{sym}$ approaches zero.
\end{proof}

\subsection{Proportional Fair Capacity}

The capacity region written in the following form is useful for
computing the proportional fair allocation:
\begin{equation}
r_i = \frac{1}{2} \log \Big( \frac{N_i + x_i P_T}{N_i + x_{i-1} P_T}
\Big) \quad \text{ for } i = 1,\ldots, K, \label{eq:alternate_form}
\end{equation}
with
\[
 0 = x_0 \leq x_1 \leq x_2 \leq \ldots \leq x_{K-1} \leq x_K = 1.
\]
The quantity $x_i P_T$ represents the sum of powers $P_1 + \ldots +
P_i$.

\begin{theorem}
Let $\Phi_i$ be a concave and monotonically increasing function for
$i=1,\ldots , K$. Using the notation in~\eqref{eq:alternate_form},
the point that maximizes
\[
 f(\vec{r}) := \sum_{i=1}^K \Phi_i(r_i)
\]
in the capacity region of a BC with total power $P_T$ and noise
powers $N_1 \leq \ldots \leq N_K$ satisfies
\begin{equation}
\frac{\Phi_i'(r_i)}{N_i + x_i P_T} =
\frac{\Phi_{i+1}'(r_{i+1})}{N_{i+1} + x_i P_T}
\end{equation}
for $i=1,\ldots, K-1$.

In particular a proportional fair allocation satisfies the equation
\begin{equation}
  r_i(N_i+x_iP_T) = r_{i+1} (N_{i+1} + x_i P_T)
  \label{eq:PF_equation}
\end{equation}
for $i=1,\ldots, K-1$. \label{theorem:broadcast_PF}
\end{theorem}

\begin{proof}
Suppose that
\begin{equation}
\vec{r} = \Big( \frac{1}{2} \log \big( \frac{N_i + x_i P_T}{N_i +
x_{i-1} P_T} \big) \Big)_{i=1,\ldots,K} \label{eq:alternate_form2}
\end{equation}
maximizes $f(\vec{r})$. This point lies on the boundary of the
capacity region. The tangent plane at $\vec{r}$ must be orthogonal
to the gradient
\[
\nabla f(\vec{r}) = \Big( \Phi_1'(r_1), \Phi_2'(r_2), \ldots,
\Phi_K'(r_k) \Big)
\]
i.e., it must be orthogonal to $\vec{r}' - \vec{r}$ for all
$\vec{r}'$ on the tangent plane.

Differentiate \eqref{eq:alternate_form2} with respect to $x_i$, we
obtain for $i=1,\ldots, K-1$,
\[
 \frac{1}{2}\Big( 0,\ldots, 0, \frac{P_T}{N_i + x_iP_T}, \frac{-P_T}{N_{i+1} + x_i P_T}, 0, \ldots, 0
 \Big).
\]
The two fractions in the above vector is in the $i$th and $(i+1)$st
component. This must be orthogonal to the gradient $\nabla
f(\vec{r})$. Equating the dot product to zero, we get
\[
\frac{1}{2} \frac{P_T}{N_i + x_iP_T} \Phi_i'(r_i) - \frac{1}{2}
\frac{P_T}{N_{i+1} + x_i P_T} \Phi_{i+1}'(r_{i+1}) = 0
\]

For the proportional fair point, we take $\Phi_i$ to be the log
function for all $i$.
\end{proof}

\begin{corollary}
Let $\vec{r}^{PF}$ and $\gamma^{PF}$ be the rate vector and SINR
vector corresponding to the proportional fair allocation, with $N_1
\leq \ldots \leq N_K$. Then the rates and SINR of the users are in
decreasing order, i.e.,
\[
 r_1^{PF} \geq r_2^{PF} \geq \ldots \geq r_K^{PF}
\]
and
\[
 \gamma_1^{PF} \geq \gamma_2^{PF} \geq \ldots \geq \gamma_K^{PF}.
\]
\label{cor:BC_PF_rate_order}
\end{corollary}

\begin{proof}
From \eqref{eq:PF_equation}, we have
\[
  r_i = r_{i+1}\frac{N_{i+1} + x_iP_T}{N_i + x_i P_T} \geq r_{i+1}.
\]
Since the rate is a monotonically increasing function of the SINR,
the inequalities about SINR follows immediately.
\end{proof}

\begin{corollary}
The infimum of $\eta_{PF}$, taken over all $K$-user Gaussian BC, is
equal to the lower bound $1/K$.
\end{corollary}

\begin{proof}
Suppose that we fix the noise powers $N_1,\ldots, N_K$ and take $P_T
\rightarrow 0$. Equation~\eqref{eq:PF_equation} implies that
\[r_i^{PF} N_i \approx r_j^{PF} N_j\] for all $i$ and $j$. When $P_T$ is small,
\[
 r_i^{PF} = \frac{1}{2} \log \Big(1+ \frac{p_i^{PF}}{N_i + p_{i-1}^{PF} + \ldots p_1^{PF}}  \Big)
 \approx \frac{1}{2} \frac{p_i^{PF}}{N_i}.
\]
Hence $p_i^{PF} \approx p_j^{PF}$ for all $i$ and $j$, and $p_i^{PF}
\rightarrow P_T/K$ at the proportional fair allocation as $P_T
\rightarrow 0$. We have the following limits,
\[
 \Csum \rightarrow \frac{1}{2} \frac{P_T}{N_1},
\]
and
\begin{align*}
\Cpf &\rightarrow \frac{1}{2} \sum_{i=1}^K \frac{p_i^{PF}}{N_i} \\
&\approx \frac{P_T}{2K}  \sum_{i=1}^K \frac{1}{N_i}.
\end{align*}
We obtain
\[
\eta_{PF} = \frac{\Cpf}{\Csum} \rightarrow \frac{1}{K} \Big( 1 +
\frac{N_1}{N_2} + \ldots + \frac{N_1}{N_K} \Big).
\]
The right hand side can be arbitrarily close to $1/K$ if $N_1\ll
N_i$ for all $i=2,3\ldots, K$.
\end{proof}

\begin{theorem} In a Gaussian BC with noise power $N_1 \leq \ldots
\leq N_K$, the corresponding powers corresponding to the
proportional fair allocation are in increasing order,
\[
 p_1^{PF} < p_2^{PF} < \ldots < p_K^{PF}.
\]
\label{theorem:BC_PF_power_order}
\end{theorem}

\begin{proof}
We first prove the theorem in a two-user case. The noise power of
user $i$ is $N_i$, ($i=1, 2$) with $N_1 \leq N_2$. By
Theorem~\ref{theorem:broadcast_PF}, the proportional fair rate
allocation is
\begin{align*}
 r_1 &= \frac{1}{2} \log \Big(\frac{N_1 + \alpha P_T}{N_1}  \Big) \\
 r_2 &= \frac{1}{2} \log \Big(\frac{N_2 +  P_T}{N_2+ \alpha P_T}
 \Big),
\end{align*}
where $\alpha$, $0\leq \alpha \leq 1$ is chosen such that
\[
 r_1 (N_1 + \alpha P_T) = r_2 (N_2 + \alpha P_T).
\]
That is, the value of $\alpha$ satisfies
\begin{equation}
(N_1 + \alpha P_T) \log \left(\frac{N_1+\alpha P_T}{N_1} \right) =
(N_2 + \alpha P_T) \log \left(\frac{N_2 + P_T}{N_2 + \alpha P_T}
\right) \label{eq:PFordering}
\end{equation}
The power of user 1 is $\alpha P_T$ and the power of user 2 is
$(1-\alpha) P_T$. We want to show that~\eqref{eq:PFordering} cannot
hold if $\alpha \geq 0.5$.

We will apply the inequality
\[
 x \log \Big( 1 + \frac{b}{x} \Big) < b
\]
which holds for  $x > \max\{ 0 , -b\}$. This inequality is an
immediate consequence of the inequality
\[
 e^x < 1+x,\ \forall x\neq 0.
\]
Applying the inequality, we get an upper bound
\[
 \text{ R.H.S. of~\eqref{eq:PFordering} } = (N_2 + \alpha P_T)
 \log\Big(1+ \frac{(1-\alpha)P_T}{N_2 + \alpha P_T} \Big)
 < (1-\alpha) P_T,
\]
and a lower bound
\[
 \text{L.H.S. of~\eqref{eq:PFordering}} = - (N_1 + \alpha P_T) \log\left(1 + \frac{-\alpha P_T}{N_1 + \alpha P_T}   \right)
 > -( -\alpha P_T) = \alpha P_T.
\]
If $\alpha \geq 0.5$, we can combine the two bounds above,
\begin{align*}
(N_1 + \alpha P_T) \log \left(\frac{N_1+\alpha P_T}{N_1} \right)
 &> \alpha P_T \geq (1-\alpha) P_T \\
 &>(N_2 + \alpha P_T) \log \left(\frac{N_2 + P_T}{N_2 + \alpha P_T}
\right).
\end{align*}
Therefore, we must have strict inequality in~\eqref{eq:PFordering}
when $\alpha \geq 0.5$. It is noted that the bounds in the above
argument is valid for any $N_1$ and $N_2$.

In a Gaussian BC with $K$ users, consider the pair of consecutive
users $i$ and~$i+1$. At the proportional fair point in the capacity
region,
\[
\Big(N_i + \sum_{j=1}^i P_i \Big) \log \left(1 + \frac{P_i}{ N_i +
\sum_{j=1}^{i-1} P_j} \right)
  =
\Big(N_{i+1} + \sum_{j=1}^i P_i \Big) \log \left(1 + \frac{P_{i+1}
}{ N_{i+1} + \sum_{j=1}^{i} P_j} \right)
\]
by Theorem~\ref{theorem:broadcast_PF}. By setting
\begin{align*}
\tilde{N}_1 &:= N_i + \sum_{j=1}^{i-1}P_i \\
\tilde{N}_2 &:= N_{i+1} + \sum_{j=1}^{i-1}P_i \\
\tilde{P}_T &:= P_i + P_{i+1}\\
\tilde{\alpha} &:= P_i/(P_i + P_{i+1}),
\end{align*}
we get
\[
(\tilde{N}_1 + \tilde{\alpha} \tilde{P}_T) \log
\left(\frac{\tilde{N}_1+ \tilde{\alpha} \tilde{P}_T}{\tilde{N}_1}
\right) = (\tilde{N}_2 + \tilde{\alpha} \tilde{P}_T) \log
\left(\frac{\tilde{N}_2 + \tilde{P}_T}{\tilde{N}_2 + \tilde{\alpha}
\tilde{P}_T} \right).
\]
We can proceed as in the 2-user case and conclude that $p_i <
p_{i+1}$.
\end{proof}

From numerical examples, we have the following conjecture.

\begin{conjecture} Let $C_{PF}(\vec{N}, P_T)$ denote the proportional
fair capacity of a Gaussian BC with noise vector $\vec{N}$ and total
power~$P_T$.
 If $\vec{N} \preceq \vec{N}'$, then $C_{PF}(\vec{N},P_T) \leq
C_{PF}(\vec{N}',P_T)$ \label{conjecture}
\end{conjecture}

\subsection{Algorithm}
We will use the notation for the capacity region
in~\eqref{eq:alternate_form}, and present an numerical algorithm
that maximizes $\sum_{i=1}^K \Phi_i(r_i)$ in the capacity region of
a Gaussian~BC. It is assumed that $\Phi_i$ is a strictly
monotonically increasing and concave function, so that the inverse
of the derivative $\Phi_i'$ is easy to compute. If $\Phi_i$ is the
logarithm function, then the resulting point is the proportional
fair solution. If $\Phi_i(x) = \log(x - d_i)$, the result is the
Nash bargaining solution with disagreement point $\vec{d}$.

Theorem~\ref{theorem:broadcast_PF} says that the equation
\[
\Phi_{i+1}'(r_{i+1}) = \Phi_i'(r_i)  \frac{N_{i+1} + x_i P_T}{N_i +
x_i P_T}
\]
must holds for the optimal solution. We use this equation to express
$x_{i+1}$ in terms of $x_i$ for $i=2,\ldots, K-1$. It reduces the
problem to a one dimensional search. Given any $x_1$ we first
compute $r_2$ by
\[
 r_2 = \Phi_2'^{-1} \Big( \Phi_1'(r_1) \frac{N_{2} + x_1 P_T}{N_1 + x_1
P_T} \Big)
\]
and get $x_2$ by solving
\[r_2 = \frac{1}{2} \log\Big( \frac{N_2 + x_2 P_T}{N_2 + x_1 P_T} \Big)
\]

In similar way, we compute $r_i$ and $x_i$ for $i=3,4,\ldots, K-1$.
Finally $r_K$ can be obtained once we know $x_{K-1}$.

Define the function $\chi(x_1)$ as
\[
  \chi(x_1) := \Phi_K'(r_K) (N_{K-1} + x_{K-1} P_T) - \Phi_{K-1}'(r_{K-1}) (N_{K} +
  x_{K-1} P_T)
\]
It is a function of $x_1$ as the variable $r_K$, $r_{K-1}$, $x_K$
and $x_{K-1}$ all depend on $x_1$. We can now search for the zero of
$\chi(x_1)$ numerically, say $\chi(x_1^*) = 0$. From $x_1^*$, we get
$x_i^*$ by method described above. The vector $(x_1^*, \ldots,
x_K^*)$ will satisfy the condition in
Theorem~\ref{theorem:broadcast_PF} and hence is the optimal
solution. Since any zeros of $\chi$ gives rise to an optimal
solution and we know that there the optimal solution is unique, the
function $\chi$ has only one zero.

\section{Conclusion}

We show how to pick a point in the capacity region of Gaussian MAC
and BC according to some fairness criteria. In the Gaussian MAC,
there is a strong notion of fairness, namely there is a point on the
dominant face that are majorized by all other points on the dominant
face, and are both max-min and proportional fair. We can thus call
this {\em the} fair point in the capacity region. In some particular
cases, the fair point can be computed in $O(K^2)$ time. For the
Gaussian BC, the problem of locating the proportional fair solution
or Nash bargaining solution reduces to a one-dimensional search. In
both channels, fair rate allocation can be compute efficiently.

\section{Appendix}
\label{sec:appendix}

\begin{proof}[Proof of Lemma~\ref{lemma:maxmin}]
($\Leftarrow$) For any $i\in \Omega$, suppose that $i$ is contained
in a bottleneck $B$, and $r_i = \max \{r_j:\, j\in B\}$. If we want
to increase $r_i$, we have to decrease $r_k$ for some other $k\in
B$. Since $r_i$ is the largest in $\{r_j:\, j\in B\}$, we must have
$r_k\leq r_i$. The vector $\vec{r}$ is thus max-min fair.

($\Rightarrow$) Conversely, suppose that $\vec{r}$ is a vector such
that $B_1, \ldots, B_T$ are all the bottlenecks that contain $i$,
and $r_i$ is not maximal in all such bottlenecks, i.e., $ r_i < \max
\{ r_j:\, j\in B_t \}$ for all $t=1,\ldots, T$. We can choose $i_t$
in $B_t$ such that $r_{i_t} > r_i$. If we increase $r_i$ by
$\epsilon$ and decrease each $r_{i_t}$ by sufficiently small
$\epsilon$, the resulting vector remains in~$\mathcal{P}(g)$. The
vector $\vec{r}$ is thereby not max-min fair.
\end{proof}

\begin{lemma}
Suppose that a function $g:2^\Omega \rightarrow \mathbb{R}_+$
satisfies the submodular property. Union and intersection of two
bottlenecks of $\vec{r}$ in $\mathcal{P}(g)$ are also bottlenecks of
$\vec{r}$. \label{lemma:polymatroid_lemma}
\end{lemma}

\begin{proof}
Suppose that $S$ and $T$ are both bottlenecks of $\vec{r}$, i.e.,
$\vec{r}(S) = g(S)$ and $\vec{r}(T) = g(S)$.
\begin{align*}
 \vec{r}(S) + \vec{r}(T) & = g(S)+g(T) \\
 &\geq g(S\cup T) + g(S\cap T) \\
 &\geq \vec{r}(S\cup T) + \vec{r}(S\cap T) \\
 &= \vec{r}(S) + \vec{r}(T)
\end{align*}
Therefore, all inequalities above are in fact equalities. In
particular, $\vec{r}(S\cup T) =g(S\cup T)$ and $\vec{r}(S\cap T)
=g(S\cap T)$.
\end{proof}

The proof of Theorem~\ref{theorem:MMPF} is divided into the next two
propositions.

\begin{proposition}
Suppose that $\vec{r}^{MM}$ be the max-min fair point
in~$\mathcal{P}(g)$, where $g$ satisfies the order property. By
relabeling, we can assume without loss of generality that
\begin{equation} r_1^{MM} = \ldots = r_{i_1}^{MM} <
r_{i_1+1}^{MM} = \ldots = r_{i_2}^{MM} < \ldots < r_{i_{L-1}+1}^{MM}
= \ldots = r_{i_L}^{MM}, \label{eq:increasing_order}
\end{equation}
where
\[
 0 < i_1 < i_2 < \ldots < i_L = K.
\]
The sets $\{ 1,\ldots, i_{\ell} \}$, $\ell = 1,\ldots, L$, are
bottlenecks of $\vec{r}^{MM}$. In particular, we have
\[
 \vec{r}^{MM}(\Omega) = g(\Omega),
\]
i.e., the max-min fair solution lies on the dominant face
of~$\mathcal{P}(g)$. \label{prop:dominant_face}
\end{proposition}

\begin{proof}
Let $B_\ell$ denote the set $\{ 1,\ldots, i_{\ell} \}$ for $\ell =
1,\ldots, L$.

For any element $j \in B_1$, there is a bottleneck $A_j$ so that
$r_j = \max \{r_i :\,i \in A_j\}$. If $j \in B_1$ and $k \not\in
B_1$, then $r_j^{MM} < r_k^{MM}$. Hence $A_j$ must be a subset
of~$B_1$, for all $j \in B_1$. By taking the union of $A_j$ over all
$j\in B_1$, we get
\[ B_1 = \bigcup_{j\in B_1} A_j,
\]
and we can conclude that $B_1$ is also a bottleneck of
$\vec{r}^{MM}$ by Lemma~\ref{lemma:polymatroid_lemma}.

By similar argument, we can show that $B_\ell$ is bottleneck of
$\vec{r}^{MM}$ for all $\ell\in \{1,\ldots, L\}$.
\end{proof}

\begin{proposition}
For $\ell=1,\ldots, L$, let $B_\ell$ be the set $\{1,2,\ldots,
i_\ell \}$, with $ 0 < i_1 < i_2 < \ldots < i_L  = K.$ Let $\vec{w}$
be a vector in $\mathbb{R}_+^K$ such that
\[
w_1 = \ldots = w_{i_1} < w_{i_1+1} = \ldots = w_{i_2} < \ldots <
w_{i_{L-1}+1} = \ldots = w_{i_L}
\]

All vectors in the region
\[
\mathcal{R} := \{ \vec{r}\in\mathbb{R}_+^K:\, \vec{r}(B_\ell) \leq
\vec{w}(B_\ell),\ \ell=1,\ldots, L \}
\]
satisfy
\[
  \sum_{j=1}^k r_{[j]} \leq   \sum_{j=1}^k w_j
\]
for $k=1,\ldots, K$. Consequently, $\vec{w}$ is majorized by all
points in $\mathcal{R}$ such that $\vec{r}(\Omega) =
\vec{w}(\Omega)$
 \label{prop:nested}
\end{proposition}

\begin{proof}
Suppose that $\vec{v}$ is a point in~$\mathcal{R}$ that does not
majorize $\vec{w}$. There is an index $k$, $1\leq k < K$, such that
\[
  \sum_{i=1}^k v_{[i]} > \sum_{i=1}^k w_i.
\]
We can find an $\ell$ so that $i_\ell \leq k < i_{\ell+1}$. (Define
$i_0 := 0$ and $B_0 := \emptyset$ if necessary.)

Consider the collection $\mathcal{T}$ of all subsets
$S\subseteq\Omega$ such that $B_\ell \subseteq S \subseteq
B_{\ell+1}$ and $|S|=k$. The number of such subsets is
\[
a := \binom{i_{\ell+1} - i_{\ell}}{k - i_{\ell}}.
\]
Since $w_i$ is constant for $i\in B_{\ell+1} \setminus B_{\ell}$, we
have
\[
  \sum_{i\in S} w_i = \sum_{i=1}^k w_i
\]
for all $S \in \mathcal{T}$. Hence
\[
 \sum_{i \in S} v_i \geq \sum_{i=1}^k v_{[i]} > \sum_{i=1}^k w_i = \sum_{i\in
 S} w_i.
\]
We sum the above over all $S\in\mathcal{T}$,
\begin{equation}
 \sum_{S\in\mathcal{T}} \sum_{i \in S} v_i  > \sum_{S\in\mathcal{T}} \sum_{i\in
 S} w_i. \label{eq:proof_of_main}
\end{equation}
The left hand side in the above inequality equals
\[
 a\sum_{i=1}^{i_\ell} v_i + b  \sum_{i=i_\ell+1}^{i_{\ell+1}} v_i,
\]
where
\[
b := \frac{a(k-i_{\ell})}{i_{\ell+1} -
 i_{\ell}}.
 \]
 Similarly, the right hand side of~\eqref{eq:proof_of_main} equals
\[
 a\sum_{i=1}^{i_\ell} w_i + b  \sum_{i=i_\ell+1}^{i_{\ell+1}} w_i.
\]
We rewrite \eqref{eq:proof_of_main} as
\[
 (a-b) \sum_{i=1}^{i_\ell} (v_i - w_i) + b \sum_{i=1}^{i_{\ell+1}} (v_i - w_i) >
 0,
\]
or equivalently
\[
 (a-b) (\vec{v}(B_\ell) - \vec{w}(B_\ell)) + b (\vec{v}(B_{\ell+1}) - \vec{w}(B_{\ell+1})) >
 0.
\]

 Since $a\geq b$, $\vec{w}(B_{\ell}) \geq \vec{v}(B_{\ell})$ and
$\vec{w}(B_{\ell+1}) \geq \vec{v}(B_{\ell+1})$, the left hand side
must be less than or equal to zero. We get a contradiction.
\end{proof}

\begin{proof}[Proof of Theorem~\ref{theorem:MAC_majorization}]
Suppose without loss of generality that $P_1 \leq \ldots \leq P_K$
and $\tilde{P}_1 \leq \ldots \leq \tilde{P}_K$. Let $\vec{r}^*$ and
$\tilde{\vec{r}}^*$ be the max-min fair point in $\mathcal{P}(g)$
and $\mathcal{P}(g')$ respectively. It is clear that $g(\{1\}) \leq
\ldots \leq g(\{K\})$, $g'(\{1\}) \leq \ldots \leq g'(\{K\})$, and
both $g$ and $g'$ satisfy the order property. By
Theorem~\ref{theorem:order}, we obtain $r^*_1 \leq \ldots \leq
r^*_K$ and $\tilde{r}^*_1 \leq \ldots \leq \tilde{r}^*_K$. We want
to show
\[
  \sum_{j=1}^k r^*_j \geq \sum_{j=1}^k \tilde{r}^*_j
\]
for $k=1,\ldots, K-1$.

As in the proof of Prop.~\ref{prop:dominant_face}, a bottleneck of
the max-min fair vector is of the form $\{1,2,\ldots \ell\}$. We can
disregard all constraints except those in the form $\vec{r}(S) \leq
g(S)$ for $S = \{1,2,\ldots, \ell\}$, $\ell=1,\ldots, K$. Let $k$ be
any integer between 1 and $K$. By Prop.~\ref{prop:nested}, the point
$\vec{r}^*$ maximizes $\sum_{j=1}^k r^*_j$ in the region

\[
\begin{cases}
r_1+\ldots+r_k \leq g(\{1,2,\ldots, k\}) &\text{ for } k=1,\ldots,
K, \\
r_k \leq r_{k+1} &\text{ for } k=1,\ldots, K-1, \\
r_k \geq 0 & \text{ for } k= 1, \ldots, K.
\end{cases}
\]

Expressed in terms of matrix, the constraints become $\vec{r}^T
\mat{A} \leq \vec{b}^T$, where
\[
\mat{A} = \left[
\begin{array}{ccccc| cccc }
1 & 1 & 1 &\cdots & 1     & 1 & & \cdots&\\
  & 1 & 1 &\cdots & 1     & -1 & 1 & \cdots&\\
  &   & 1 &\cdots & 1     & & -1& \cdots &\\
  &   &   &       & \vdots && & \cdots &1\\
  &  &    &        & 1    & & & \cdots & -1
\end{array} \right]\]
and
\[
 \vec{b}^T = [g(\{1\}), g(\{1,2\}), \ldots, g(\{1,2,\ldots, K\}), 0 ,\ldots, 0 ].
\]
Let $\vec{c}$ denote the column vector $[1, \ldots, 1, 0, \ldots,
0]^T$ in which exactly $k$ components are equal to 1. The maximal
value in the linear program
\begin{gather*}
 \max\ \sum_{j=1}^k r_j = \vec{c}^T \cdot \vec{r} ,\\
 \vec{r}^T \mat{A} \leq \vec{b}^T \\
 \vec{r} \geq \vec{0}
\end{gather*}
is equal to $\sum_{j=1}^k r^*_j$. By duality of linear programming,
the minimal value of the dual problem, with $s_j$'s as the dual
variables,
\begin{gather*}
\min \sum_{j=1}^K g(\{1,\ldots, j\}) s_j, \\
\mat{A} \vec{s} \geq \vec{c}
\\
\vec{s} \geq \vec{0}
\end{gather*}
coincides with $\sum_{j=1}^k r^*_j$.

Since $\vec{P} \preceq \vec{P}'$, we know that $g(\{1,\ldots, j\})
\geq g'(\{1,\ldots, j\})$ for $j=1,\ldots, K$. If we replace
$g(\{1,\ldots, j\})$ by $g'(\{1,\ldots, j\})$ , we obtain another
linear program,
\begin{gather*}
\min \sum_{j=1}^K g'(\{1,\ldots, j\}) s_j, \\
\mat{A} \vec{s} \geq \vec{c}
\\
\vec{s} \geq \vec{0}.
\end{gather*}
The minimal value equals $\sum_{j=1}^k \tilde{r}^*_j$. We are
optimizing a smaller objective function over the same feasible
region. As a result, we must have $\sum_{j=1}^k r_j \geq
\sum_{j=1}^k \tilde{r}^*_j$.
\end{proof}

\vspace{1cm}

{\bf Acknowledgements:} We would like to thank Michael Ng for his
valuable discussions.

\bigskip



\end{document}